\documentclass{article}

\usepackage{arxiv}

\usepackage[utf8]{inputenc} 
\usepackage[T1]{fontenc}    
\usepackage{hyperref}       
\usepackage{url}            
\usepackage{booktabs}       
\usepackage{amsfonts}       
\usepackage{nicefrac}       
\usepackage{microtype}      
\usepackage{lipsum}
\usepackage{graphicx}
\graphicspath{ {./images/} }

\usepackage{comment}
\usepackage{amsmath}
\usepackage{amsfonts}
\usepackage{booktabs}
\usepackage{algorithm}
\usepackage{algorithmic}
\usepackage{color}
\usepackage{mathtools}
\usepackage{mathrsfs}
\usepackage{subfigure}

\newcommand{\new}[1]{#1}
\newcommand{\newedit}[1]{#1}

\title{Interaction Mix and Match: Synthesizing Close Interaction using Conditional Hierarchical GAN with Multi-Hot Class Embedding}

\author{
 Aman Goel \\
  International Institute of Information Technology\\
  Hyderabad, India \\
  \texttt{aman.goel@students.iiit.ac.in} \\
   \And
 Qianhui Men \\
 Department of Engineering Science\\
 University of Oxford, United Kingdom \\
  \texttt{qianhui.men@eng.ox.ac.uk} \\
  \And
 Edmond S. L. Ho\thanks{corresponding author} \\
  School of Computing Science\\ University of Glasgow, United Kingdom \\
  \texttt{Shu-Lim.Ho@glasgow.ac.uk} \\
}

\begin{document}
\maketitle
\begin{abstract}
Synthesizing multi-character interactions is a challenging task due to the complex and varied interactions between the characters. In particular, precise spatiotemporal alignment between characters is required in generating close interactions such as dancing and fighting. Existing work in generating multi-character interactions focuses on generating a single type of reactive motion for a given sequence which results in a lack of variety of the resultant motions. In this paper, we propose a novel way to create realistic human reactive motions which are not presented in the given dataset by mixing and matching different types of close interactions. We propose a Conditional Hierarchical Generative Adversarial Network with Multi-Hot Class Embedding to generate the Mix and Match reactive motions of the follower from a given motion sequence of the leader. Experiments are conducted on both noisy (depth-based) and high-quality (MoCap-based) interaction datasets. The quantitative and qualitative results show that our approach outperforms the state-of-the-art methods on the given datasets. We also provide an augmented dataset with realistic reactive motions to stimulate future research in this area. The code is available at \url{https://github.com/Aman-Goel1/IMM}
\end{abstract}


\begin{figure*}[b]
   \begin{flushright}
   \includegraphics[width=\linewidth]{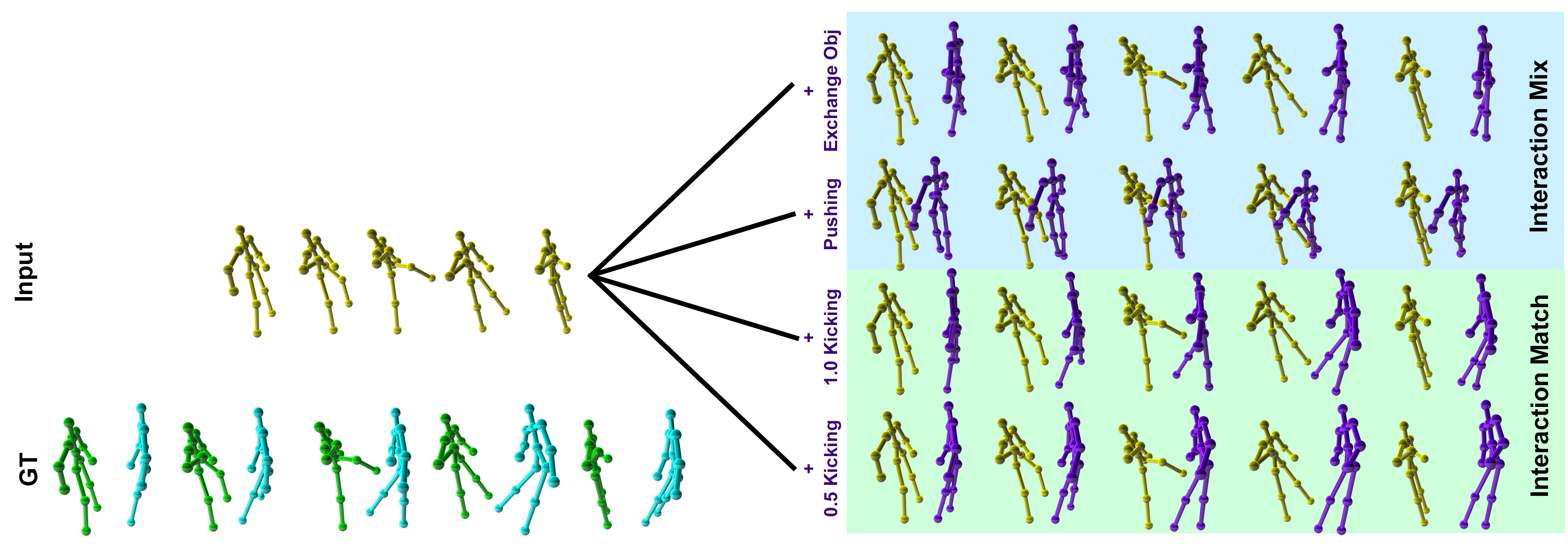}
   \end{flushright}
   \caption{Examples of Mix And Match Interactions. Given the input motion (yellow) of one character, different reactive motions (purple) can be generated by specifying the interaction labels. }
   \label{fig:mix_n_match}
\end{figure*}

\section{Introduction}

Animating virtual human-like characters has been playing an important role in a wide range of applications, including computer games, CGI/3D anime movies, virtual reality, etc. Character animation was traditionally a labour-intensive task as it requires extensive manual intervention from experienced animators to generate high-quality animations. With the advancement of motion acquisition technology, pre-recorded human motions are more accessible and can be used for animating virtual characters to semi-automate the animation production pipeline. However, a significant amount of manual intervention and post-processing is still required to remove artefacts caused by noise, marker swaps, and marker occlusions in the captured motions \cite{Perepichka:MIG19}.

Although encouraging results are demonstrated by using data-driven and deep learning techniques \cite{Holden:DL16,Ling:MotionVAE,Starke:Sig21} in animating virtual characters in recent years, most of the existing work focuses on generating single-character or character-object interactions. On the other hand, generating multi-character (two or more characters) interactions is less researched in the literature. Synthesizing close interactions, such as dancing and hugging, between multi-character is a challenging research problem since the motions of the characters have to be precisely aligned spatially and temporally to avoid artefacts such as interpenetration of body parts while preserving the contextual meaning in the interaction (such as the contact at the hands in high-five). To better support animators in generating a wide variety of two-character interactions efficient with a high degree of controllability, we propose a method for \textit{Interaction Mix and Match} in this research. Specifically, given the motion of a single character as an input, our method enables the synthesis of \new{1) \textit{diverse reactive motions} by adjusting the scale of input label from the training data, 
and 2) \textit{new interactions types} by mixing the interaction types informed by the multi-class label embeddings.}

An intuitive solution will be combining individually captured motions \cite{Shum:VRST07,Shum:SA08,Komura:CAVW05} and editing the interaction according to additional constraints given by the users. However, careful design of the motion editing algorithms (such as \cite{Ho:ToG10,Ho:EG09}) and parameter tuning are required to avoid interpenetration of body parts. Synthesizing a virtual partner/opponent from the user's movement has been explored in VR dancing \cite{Ho:TOMM13} and sword fighting \cite{Dehesa:CHI20}. 
\newedit{While the aforementioned approaches can generate the reactive motion (i.e. motion reacts to the human user) of the virtual character interactively, the reaction is highly similar to the pre-recorded motions in the dataset and results in a lack of variety in the synthesized interactions.} On the other hand, modelling two-character interaction using Recurrent Neural Networks (RNNs) \cite{Kundu:WACV20} have demonstrated encouraging results in predicting the future movements. The key insight is to model the correlation between the motions of the two characters by a 2-stream cross-conditioned network \cite{guo2021multi}. Although these approaches model interactions effectively for recognition and prediction tasks, they are less desirable for animation synthesis due to the low controllability of the resultant motions. 
\new{Aristidou~et~al.~\cite{Aristidou:TVCG2022} recently proposed a music-driven approach for synthesizing dancing motion including partnered dance (e.g. Salsa). Although this work shares similar interests as ours in generating close interactions with high-level control, \cite{Aristidou:TVCG2022} is specifically designed for dancing while our proposed method can be applied to synthesizing different kinds of leader-follower interactions.}

The most relevant research to our work is the GAN-based interaction synthesis framework proposed by Men et al. \cite{Men:CG2022}. 
\new{To the best of our knowledge, \cite{Men:CG2022} is the most recent method that generates the reactive motion of the follower according to the input motion of the leader.} 
\new{\cite{Men:CG2022} considered a seq2seq reactive motion generator and adversarially rectify the synthesized motion with a binary and a multi-class discriminator. However, with no label information guided, their model can only produce a fixed reactive pattern learned from the input motion data and fails to create a mixture of reactions.}

In this paper, we introduce a novel concept of using multi-hot class embedding in a conditional GAN to generate higher quality reactive motions with a larger degree of controllability over recent research \cite{Men:CG2022}. The main goal of this work is to enable \textit{Interaction Mix and Match} which is illustrated in Figure \ref{fig:mix_n_match}. \new{For \textit{interaction mix}, we aim to create reactions that combine different types of reactive styles directed by the multi-label indicator. By modifying the multi-hot class embedding, we can synthesize different reactive motions which are not presented in the motion datasets. For \textit{interaction match}, we aim to create a single type of reactive motion in response to the input motion of the other character. By adjusting the numeric scale of the label indicator, our trained reaction generator can also create different levels of reactive variations.}


Experimental results indicate that the interactions synthesized by our proposed method outperformed the state-of-the-art reaction synthesis model \cite{Men:CG2022} as well as the baselines qualitatively and quantitatively. Numerically, lower \newedit{Average Frame Distance (AFD) and Fréchet Inception Distance (FID)} were obtained using our method which indicates our synthesized motions better resemble the original data. Qualitatively, our generated reaction has better synthesis quality with natural movements and flexible reactive styles. The synthesized reactions also show an improved representation space with clearer classification boundaries compared with non-label guided generations. We further demonstrate the positive impact of using our synthesized motions to improve the robustness of interaction recognition models through data augmentation.

\subsection{Contributions}
The contributions of this work can be summarized as follows:
\begin{itemize}
    \item We proposed a new framework for generating diverse reaction given an input action using Conditional Hierarchical GAN
    \item We proposed using Multi-Hot Class Embedding to enable users to specify the action class to enhance controllability
    \item A new synthetic 2-character close interactions dataset will be available to stimulate the research in this area
\end{itemize}

\section{Related Work}
In this section, we will first review the related research in synthesizing multi-person interactions, which is roughly divided into algorithmic (Section \ref{sec:algor_IS}) and data-driven (Section \ref{sec:data_IS}) approaches. The action-based motion synthesis approaches will then be reviewed in Section \ref{sec:action_IS}.

\subsection{Algorithmic Interaction Synthesis} \label{sec:algor_IS}
Synthesizing close interactions with two or multiple characters such as dancing and wrestling has been a challenging task in character animation. Early work in this area combines individually captured kickboxing motions to create two-character fighting scenes by constructing an action level motion graph \cite{Shum:VRST07}. By further obtaining the potential future actions between the characters through game tree expansion, the best actions are selected by the min-max algorithm. Such close interaction patterns can be pre-computed and stored as \textit{Interaction Patches} \cite{Shum:SA08} for synthesizing new multi-character scenes by concatenating different patches. To simulate hit-and-react interactions, momentum-based inverse kinematics \cite{Komura:CAVW05} is proposed to edit pre-recorded motions according to the strength of the external perturbation and the point of contact on the body.

Another stream of research mainly focuses on synthesizing motions, such as wrestling, which are difficult to be captured even if the motions are captured individually. Ho and Komura \cite{Ho:PG07} proposed using \textit{tangle} and Gauss Linking Integral (GLI) to model the entangling body parts in close interactions. Such a topologically-based pose representation can be used for synthesizing tangling or detangling body parts by increasing or decreasing the magnitude of the GLI value accordingly. The topological approach is further combined with Rapidly-exploring random trees (RRT) \cite{Ho:Humanoid07} to synthesize interactions such as carrying and piggybacking, as well as extending to \textit{Topology Coordinates} \cite{Ho:EG09} for synthesizing human-human and human-object close interactions by linearly interpolating key poses in the topology-based coordinate system.

To further provide animators with the flexibility to generate a wide range of close interactions, \textit{Interaction Mesh} \cite{Ho:ToG10} is proposed to preserve the spatial relations (i.e relative distances) between the character and its surroundings (including other characters and objects \cite{Ho:ICRA13}) for motion adaption and motion retargeting. A volumetric mesh is constructed by applying Delaunay Tetrahedralization on a 3D point cloud sampled from the key locations on the character(s) and the object(s). By minimizing the deformation of the mesh during motion adaptation, the spatial relations between the character(s) and objects can be maintained. \textit{Aura Mesh} \cite{Jin:CFG18} is proposed to capture and preserve the spatial relations at skin-level when retargeting close interactions. Naghizadeh and Cosker \cite{Naghizadeh:2019} proposed giving higher priority to local connections over global ones in Interaction Mesh to enable large-scale transformation in multi-character motion retargeting. Kim et al.~\cite{Kim:CAVW21} recently proposed using As-Rigid-As-Possible (ARAP) deformation on the pose graph for retargeting multiple characters. By avoiding computationally expensive spacetime optimization, multi-character motions can be retargeted interactively.

While the aforementioned approaches can effectively synthesize and edit close interactions, careful design of the interaction representations and parameter tuning are required which result in difficulties in generating large-scale close interactions efficiently.

\subsection{Data-driven Interaction Synthesis} \label{sec:data_IS}
With the availability of interaction datasets \cite{Yun:SBU, Shen:TVCG2020,guo2021multi,Coppola:ROMAN17}, data-driven approaches are becoming more popular. By retrieving pre-recorded interactions, virtual partner/opponent can be synthesized based on the user's motion in dancing \cite{Ho:TOMM13} and sword fighting \cite{Dehesa:CHI20} in VR. Kundu et al.~\cite{Kundu:WACV20} proposed using Cross-Conditioned Recurrent Networks for synthesizing human-human interactions. Specifically, given the observations (i.e pose sequence) of each character, two recurrent neural networks are used for predicting the motion of a character using another character's motion as input. Wen et al.~\cite{guo2021multi} also proposed a 2-stream network with cross-interaction attention (XIA) module for predicting the motion based on the previous movements of the interacting characters. Huang et al.~\cite{huang2015approximate} generated reactive motion with maximum-entropy inverse optimal control (ME-IOC). However, their model can only sample the reaction from the training dataset. The most relevant work is the interaction synthesis framework proposed by Men et al.~\cite{Men:CG2022} in which a GAN-based model is used for generating the reactive motion of a character in response to the motion of another character given as input. \newedit{By having an addition discriminator to predict the interaction class, the GAN-based model generates  motions with better quality by taking into account the class information.} 

Data-driven approaches showed promising results in synthesizing high-quality interaction with minimal human intervention. However, the existing work lacks the controllability required by animators. Our proposed method addresses this problem by proving users with a high-level control (i.e. action label) to generate the desired close interactions easily.

\subsection{Action-level Motion Synthesis} \label{sec:action_IS}

Synthesizing animation based on high-level controls such as `walk', `run' and `jump' has been an active research area in character animation. Such an intuitive control is similar to instructing actors and actresses by the director in the real world. Arikan et al.~\cite{Arikan:Sig03} proposed an optimization-based method to select relevant poses from the database to construct the resultant motion according to the input annotations (i.e. actions) while satisfying spatiotemporal constraints (such as reaching a particular location at a particular frame). A recent work proposed by Lee et al.~\cite{Lee:Sig21} focuses on using reinforcement learning to synthesize time-critical motions from interactive character control including high-level action labels. Specifically, the teacher policy is first learned to achieve the tasks in optimal ways and then the time-critical student policy is followed to improve the responsiveness of the interactive control. Battan et al.~\cite{Battan_2021_WACV} synthesize motion from the input label and give an initial set of frames using a 2-stage approach on an encoder-decoder architecture. In particular, the first stage predicts a sparse set of keyframes for the whole motion while the second stage generates the dense motion trajectories from the output of the first stage.

Action-conditional generative models have also formed a popular stream for motion synthesis in recent years. Guo et al.~\cite{guo:MM20} proposed a Lie Algebra
based Variational Auto-Encoder (VAE) framework to generate motions according to the input action label. Petrovich et al.~\cite{petrovich21actor} proposed a Transformer-based conditional VAE for 3D motion synthesis. The authors further demonstrated the effectiveness of using the learned sequence-level latent space for denoising noisy input such as the 3D pose sequence estimated from monocular video. 

While the aforementioned action-conditioned approaches have been widely used in synthesizing single-person motions, less attention has been paid to synthesizing multi-person interactions using high-level controls. The recent MUGL~\cite{Maheshwari:WACV22} adopted the Gaussian Mixture VAE (GMVAE)~\cite{Dilokthanakul:arXiv16} which can generate single and multi-person 3D motions directly from an action label. However, the method does not 1) generate the reactive motion according to the input motion and 2) support the generation of motion from multi-class labels to further increase the diversity.




\section{Methodology}
\begin{figure*}[htb]
  \centering
\includegraphics[width=0.9\linewidth]{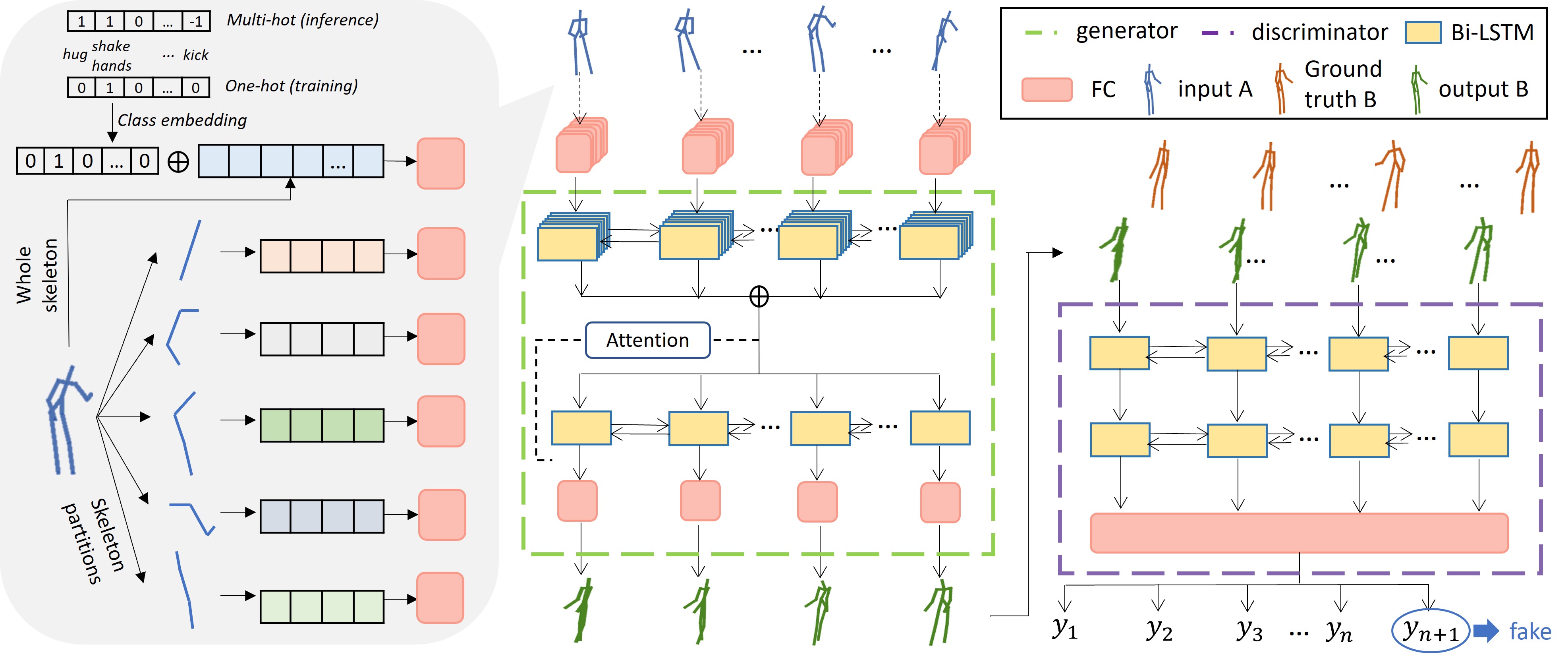}
  \caption{\new{The overview of the proposed framework. The call-out box on the left shows the label embedding and the body partition embedding of the input character. We then illustrate the attentive-based seq2seq architecture of our reactive motion generator in the middle. On the right hand side, the multi-class discriminator formed by Bi-LSTM layers is used for improving the quality of the synthesized reaction.}}
  \label{fig:overview}
\end{figure*}

In this section, we propose an end-to-end framework to synthesize stylized reactive motion informed by multi-hot action labelling. The goal of our model is to synthesize reactive patterns given an input action and its label indicator. For \textit{interaction mix}, we aim to generate reactions that combine different classes of reactive styles directed by the multi-label indicator. For example, when given an input action of \textit{kicking}, the user can specify the reactive motion to be \textit{hugging} while \textit{avoiding}. For \textit{interaction match}, we aim to generate 
the reactive motion corresponding to the interaction type and motion of the input. With the generative nature of our model, diverse interaction variations can be created. 
The overview of the proposed framework can be found in Figure \ref{fig:overview}. The generator is formed by a seq2seq attentive network with the label embedding that learns the class-specific patterns, thus creating the controllable reactions when given the multi-hot labelling during inference. The multi-class discriminator with multi-layer sequential encoding is to generate high-quality reactive patterns and improves the representation space for reaction extrapolation.

\subsection{Class Embedding}
We propose a multi-hot class indicator to control the pattern of the generated reactive motion. Class encoding, which is usually modelled as a one-hot vector, is frequently adopted in conditional GAN-based models~\cite{mirza2014conditional} to control the generator that only generates a certain class of samples. Here, we extend the one-hot encoding to multi-hot scenarios that softened the constrain to be multi-labelled. \new{More specifically, for example, ``1" is a positive label that the reactive motion is expected to perform, ``0" is a neutral label while ``-1" is a negative label that avoids generating the corresponding reaction.} Theoretically, the multi-hot labelling can also be extended to floating-point numbers to show the continuous effect of content change in the generated motion. With the Multi-Hot Class Embedding, users can specify the reactive motion that combines the characteristics of different reactive styles.

Given an input action for character A denoted as $\mathbf{X}^{A}=\{x^A_t\}_{t=1}^T$, we separately model its skeleton hierarchy to encode the spatial features. To better analyze the input action, we consider a hierarchical encoder to model the spatial dynamics for each of the five body parts as well as the whole body\new{, which results in six body structures, as illustrated in Figure~\ref{fig:overview}}. This is because modelling the actions within a group of joints can better learn the spatial correlations than modelling the whole body structure at once. For example, arm motions are more informative than other body parts in the upper body movements in interactions such as \textit{shaking hands}. 
Instead of attaching the label information to each body slice, we only concatenate with the label embedding with the whole skeleton. This is to avoid the abuse of class labelling which may lead to an over-fitted generation. The one-hot labelling during training is defined by:
\begin{equation}
1_{C}(x)=\left\{\begin{array}{ll}
1 & x\in C \\ 0 & x\notin C
\end{array}\right.
\end{equation}
Here, $C$ is the corresponding class label for the input action $\mathbf{X}^{A}$. Before feeding to the generator, each of the five body part and the label-concatenated whole body skeleton is further encoded by a fully-connected layer respectively to encode the structural feature. 

\newedit{During inference, the interaction label can either be one-hot encoding or multi-hot encoding to realize \textit{interaction mix} by combining different types of reactive motion. An example is shown in the upper-left corner of Figure \ref{fig:overview}.} The multi-hot inference with one-hot training also ensures an extrapolative representation space (as shown in Figure~\ref{fig:t-SNE}) that a multi-pattern reactive motion can be blended to satisfy the specialized reactive conditions.

\subsection{Generator Backbone}
To generate realistic reactive motion, we utilize the conditional GAN with a multi-class discriminator to improve the quality of the synthesized class-informed reaction. Men et al.~\cite{Men:CG2022} designed two discriminators of a binary and a multi-class discriminator that cooperate to create realistic motion reactive to the input motion. In contrast, we discard the binary discriminator to distinguish real or fake generated motion to better train the GAN network with high variation in the generative model, which helps alleviate the mode collapse~\cite{salimans2016improved} so that the reactive motions are less likely to be reduced to the same patterns.   

Our generator $G$ is formed by an attentive Long Short-Term Memory (LSTM)-based sequence-to-sequence (seq2seq) architecture. The encoder consists of one-layer bidirectional LSTM (Bi-LSTM)~\cite{zhang2015bidirectional} for each of the six body structures to model the temporal dynamics. 
\newedit{The six output hidden states of the encoder are concatenated and fed into the decoder which is formed by a single-layer bidirectional LSTM to create the reactive motion.} A frame-level attention mechanism is further introduced to enhance the connectivity between the encoder and the decoder, such that the model can recognize the most informative frames from input A and present similar importance to the reactive B. \newedit{Assume that the concatenated hidden state of the encoder at frame $t$ is $h_t$ and $\hat{h}_t$ is the hidden state for the decoder, the seq2seq attention learns a contextual embedding $c_t$ for decoder that allocates information from all steps in the encoded states:}:
\begin{equation}
c_{t}=\sum_{t_e=1}^{T} \phi(t_e,t)h_{t}.\label{eq:context}
\end{equation}
where $\phi(t_e,t)$ is the attention weight that evaluates the correspondence between the state $h_{t_e}$ at current encoder step $t_e$ and the previous decoder state $\hat{h}_{t-1}$:
\begin{equation}
\phi(t_e,t)=\sigma_{1}(W_{1}\sigma_{2}(W_{2}[h_{t_e};\hat{h}_{t-1}])),\label{eq:addressing}
\end{equation}
where $W_{1}$ and $W_{2}$ are the learnable parameters to increase capability of the seq2seq attention, and $\sigma_{1}$ and $\sigma{2}$ are \textit{softmax} and \textit{tanh} activation functions, respectively. Here, we consider a global seq2seq attention where all timestamps of the encoder are included. However, it is also feasible to a local embedding where only partial input motion is observed for online predicting. The contextual embedding $c_{t}$ is updated along with the hidden state at every time step, which sets up a prompt reaction by assigning step-wise significance to the decoder. Since the performance length of the input action of character A and the generated reaction for character B should be equal as in real-world interaction, $T_e$ is the same as $T$ in our seq2seq model. Furthermore, we connect a fully-connected layer to the end of the attentive Bi-LSTM decoder to reconstruct the reactive pose at every frame $t$.

\subsection{The Multi-class Discriminator}
\label{sec:3.2}
We propose a multi-class discriminator that recognizes which interaction type the synthesized reactive motion $\hat{x}_B$ belongs to. The class-wise discriminator can help increase diversity in the reactive patterns to cater for different types of input action. The architecture of the multi-class discriminator is presented on the right hand side of Figure~\ref{fig:overview}. Besides the $n$ classes of the reactive labels $y_i, i=1,...,n$, we also include an extra fidelity label $y_{n+1}$ to \new{compensate for the functionality of a binary discriminator. To this end, the reaction to be judged will belong to either one of the real types of interaction class $y_{i}$ or a fake class $y_{n+1}$.}

Instead of feeding in the interaction with motions from both characters A and B, the discriminator only recognizes the reactive motion for B to justify the discrimination of different reactive patterns. It reduces the dependency for the discriminator on the extracted features from input action A and avoids generating collapsed results~\cite{che2016mode,salimans2016improved}.

Specifically, our multi-class discriminator $D$ consists of two-layer bidirectional LSTMs to classify the synthesized reactive motion from both the forward and backward movements. Compared to a single-directional LSTM, a discriminator using Bi-LSTM can extract high-level semantic features that significantly improve the performance for sequential classification. The hidden output at every time step is concatenated and embedded into a fully-connected layer with \textit{softmax} activation to output the class probability.

\subsection{Loss Function}
During training, the synthesized reactive motion corresponding to B's motion $\mathbf{\hat X}^{B}$ is expected to be as close as possible to $\mathbf{X}^{B}$ with the $L_1$ norm to contrast their intensity similarities. The generated $\mathbf{\hat X}^{B}$ is also expected to be classified as the same class of $\mathbf{X}^{A}$. The corresponding adversarial loss for the proposed conditional GAN is defined as:
\begin{equation}
\begin{split}
\mathcal{L}_{CGAN}=&-\mathbb{E}_{x,y\sim G}\log p(y|x,y<N+1)\\&+\mathbb{E}_{x,y\sim G}\log p(y|x,y=N+1)\\&+\mathbb{E}_{x,y\sim p_{B}}\log p(y|x,y<N+1),\label{eq:l_cgan}
\end{split}
\end{equation}
where $p_{B}$ represents the real distributions of the character B's motion, $y$ is the class label, and $p(y|x)$ stands for the probability of $x$ being recognized as the synthesized interaction class. 

Besides the $L_1$ and $L_{CGAN}$ loss, we also adopt the continuity loss $L_c$ and bone loss $L_b$ introduced from~\cite{yan2017skeleton} and~\cite{Men:CG2022} to preserve the motion consistency and the bone length nature. \new{Here, we set the weight of the continuity loss to $1$ instead of $0.01$ used in \cite{Men:CG2022} and resulted in better motion quality as demonstrated in the accompanying video demos. } The overall loss function follows a min-max optimization scheme:
\begin{equation}
\min\limits_{G}\max\limits_{D}\mathcal{L}_{CGAN}+\lambda_{b}\mathcal{L}_{b}+\lambda_{c}\mathcal{L}_{c}+\lambda_{1}\mathcal{L}_{1},\label{eq:overall}
\end{equation}

\section{Experimental Settings}
In this section, the settings of the experiments will be presented. We first introduce the datasets used in this work in Section \ref{sec:datasets}. The corresponding experimental protocols, such as data split, are explained in Section \ref{sec:datasets_split}. Finally, the implementation details will be given in Section \ref{sec:imp_details}. 

\subsection{Datasets} \label{sec:datasets}
\subsubsection{SBU Two-person Interaction Dataset}
The SBU dataset \cite{Yun:SBU} contains eight classes of two-person interactions, including \textit{approaching, departing, pushing, kicking, punching, exchanging objects, hugging,} and \textit{shaking hands}. The \textit{approaching} and \textit{departing} interactions are excluded in our experiments since their reactive motions are standing still without meaningful movements. The interactions were captured using the Microsoft Kinect (depth-based sensor) and the 3D skeletal motions are provided. There are 7 subjects participated in the data collection with 197 motion sequences being used in total, and each character is represented by the locations of 15 joints in each frame.


\subsubsection{Character-Character (2C) Dataset}
The 2C dataset \cite{Shen:TVCG2020} contains high-quality kickboxing motions which are captured using an optical MoCap system. It contains 2 classes of two-person interactions, i.e., \textit{kicking} and \textit{punching} with diverse reactive patterns such as avoiding or being hit. 44 motions are used in the model and each character is represented by 20 joints in each frame. The joint information of the two characters is represented by 3D angular values with a skeleton hierarchy.

\subsection{Dataset Settings} \label{sec:datasets_split}
Same as~\cite{Men:CG2022}, we perform leave-on-subject-out cross-validation on the SBU dataset. For 2C, we pre-process it by converting the joint angles into joint positions with forward kinematics (FK). We also normalize its skeleton by a scaling factor of 100 for both training and evaluation. The train:test ratio in 2C is 3:1. The 3D joint positions for both datasets are made relative to the root joint of character A, and the global root translation and rotation about the upward vector (i.e. y-axis) at the root joint of character A are removed as data standardization.

\subsection{Implementation Details} \label{sec:imp_details}
The code base is built upon Keras platform, running on a PC with AMD Ryzen 7 3700x 8-Core Processor 4.4GHz, 64GB Memory and Nvidia RTX 2080 Ti Graphics card. RMSprop with a learning rate of 0.01 is used as the optimizer. There are 80 and 200 LSTM Neurons for each spatial slice and 480 and 1200 for the attentive layer for SBU and 2C, respectively. A batch size of 16 was used for both datasets, and we train 1600 epochs and 2000 epochs on SBU and 2C, respectively. For the weights of the network Loss, we set \begin{math} \lambda_{b} \end{math} = 0.01, \begin{math} \lambda_{c} = 1 \end{math}, \begin{math} \lambda_{1} \end{math} = 1.
Since we emphasize the continuity loss to create motions of better quality, a higher weight is given to it to encourage more smooth joint trajectories to be produced.

\section{Experimental Results}
Extensive experiments have been conducted to evaluate the effectiveness of our proposed method. Firstly, we evaluate the quality of the motion synthesized by our method quantitatively and qualitatively in Section \ref{sec:exp_motionSyn}. Secondly, we visualize the learned latent space in Section \ref{sec:exp_latent} to demonstrate how our method facilitates the synthesis of two-character interactions with high-level controls. Thirdly, we justify the design of our proposed framework by conducting an ablation study in Section \ref{sec:exp_ablation}. Finally, we demonstrate another application of our proposed method as a data augmentation approach to improving the robustness of interaction recognition models by providing a wider variety of synthesized training data in Section \ref{sec:recognitoin}. \new{We also compare our results with those obtained using the implementation of the state-of-the-art reaction synthesis model \cite{Men:CG2022} provided by the authors to highlight the superior performance of our method.}

\subsection{Motion Synthesis} \label{sec:exp_motionSyn}
In this section, the quantitative and qualitative evaluations of the results generated by our methods and the baselines are presented in the following subsections.

\subsubsection{Quantitative Analysis}
For quantitative evaluation, we adopted the deterministic metric Average Frame Distance (AFD) to compare the synthesized skeletal motion with the ground truth in terms of geometric similarity:
\begin{equation}
    AFD:=\frac{1}{T}\sum^{t}||x_{t}^{B'} - x_{t}^{B}||
\end{equation}
where $x_{t}^{B'}$ and $x_{t}^{B}$ are the synthesized and ground truth skeletal pose at time $t$, respectively, and $1 \leq t \leq T$. The AFD on different interaction classes on the SBU dataset is shown in Table \ref{tab:SBU_AFD}. We compare our model with the traditional machine learning methods including Nearest Neighbour (NN), Hidden Mixture Model (HMM), Discrete Markov Decision Process~\cite{kitani2012activity}, kernel-based reinforcement learning~\cite{huang2014action}, maximum-entropy inverse optimal control~\cite{huang2015approximate}, and the recent deep-learning based method~\cite{Men:CG2022}. The results show that our method outperformed all existing methods by achieving the lowest AFD in all 6 classes. Furthermore, our results show a more consistent  performance across different class with a small range of AFD (0.32-0.50) while a large range of AFD (0.44-0.72) is obtained by the state of the art \cite{Men:CG2022}.

\begin{table}[htb]
    \centering
    \resizebox{\columnwidth}{!}{%
\begin{tabular}{|l|c|c|c|c|c|c|c|} 
 \hline
 & \multicolumn{7}{c|}{AFD (↓)} \\
 \hline
 Action & NN & HMM & \cite{kitani2012activity} & \cite{huang2014action} & \cite{huang2015approximate} & \cite{Men:CG2022} & OURS\\
 \hline
Kick  & 0.81 & 0.92 & 0.65 & 0.92 & 0.67  &0.53&    \textbf{0.50}\\
Push&  0.51  & 0.60 & 0.45 & 0.61 & 0.48 & 0.52   & \textbf{0.43}\\
 Shake hand& 0.48 & 1.41 & 0.42 & 0.54 & 0.42  & 0.44 &\textbf{0.40}\\
 Hug    &0.61 & 0.67 & 0.48 & 0.81 & 0.47 & 0.72& \textbf{0.42} \\
 Ex. obj.& 0.63 & 3.84 & 0.53 & 0.74 & 0.54 & 0.45  &\textbf{0.40}\\
  Punch &0.56 & 0.66 & 0.48 & 0.66 & 0.52 & 0.45&   \textbf{0.32}\\
 \hline
\end{tabular}%
}
    \caption{The AFD of different actions in the SBU dataset. \new{Our method achieves the lowest AFD for all types of interactions.}}
    \label{tab:SBU_AFD}
\end{table}

In addition to AFD, we further compute the Fréchet Inception Distance (FID) to measure the difference in the distribution between the synthesized and original motions in the SBU and 2C datasets. The results are presented in Table \ref{tab:FID_SBU} and \ref{tab:FID_2C}. Again, our method outperformed \cite{Men:CG2022} by archiving a lower FID in all classes on both the SBU and 2C datasets. This indicates the motions synthesized using our method can consistently better resemble the motion quality as in the original motions in different datasets.

\begin{table}[htb]
    \centering
    \resizebox{\columnwidth}{!}{%
\begin{tabular}{|l|c|c|c|c|c|c| } 
 \hline
 & \multicolumn{6}{c|}{FID (↓)} \\
 \hline
  Method & Kick & Push & Shake hand & Hug & Ex. obj. & Punch \\
 \hline
 \cite{Men:CG2022} & 10.8   & 20.8 & 23.8 & 29.5 & 16.7 & 11.2\\
 OURS   & \textbf{9.3}   & \textbf{16.8} & \textbf{15.1} & \textbf{19.3} & \textbf{11.7} & \textbf{10.8}\\ 

 \hline
\end{tabular}%
}
    \caption{The FID of different actions in the SBU dataset. \new{Our method achieves a lower FID in all interaction classes.}}
    \label{tab:FID_SBU}
\end{table}

\begin{table}[htb]
    \centering
\begin{tabular}{|l|c|c| } 
 \hline
 & \multicolumn{2}{c|}{FID (↓)} \\
 \hline
 Method & Kick & Punch \\
 \hline
 \cite{Men:CG2022} & 194.2   & 148.1\\
 OURS   & \textbf{164.2}   & \textbf{122.4}\\ 

 \hline
\end{tabular}
    \caption{The FID of different actions in the 2C dataset. \new{Our method achieves a lower FID in both kick and punch classes.} }
    \label{tab:FID_2C}
\end{table}

\subsubsection{Qualitative Analysis}
To assess the visual quality of the interactions synthesized by our method, readers are referred to the accompanying video demo. In this section, we will qualitatively evaluate the synthesized motions in different experimental settings. 

\begin{figure}[htb]
\centering
    \subfigure[Kick]{
        \includegraphics[width=0.8\linewidth]{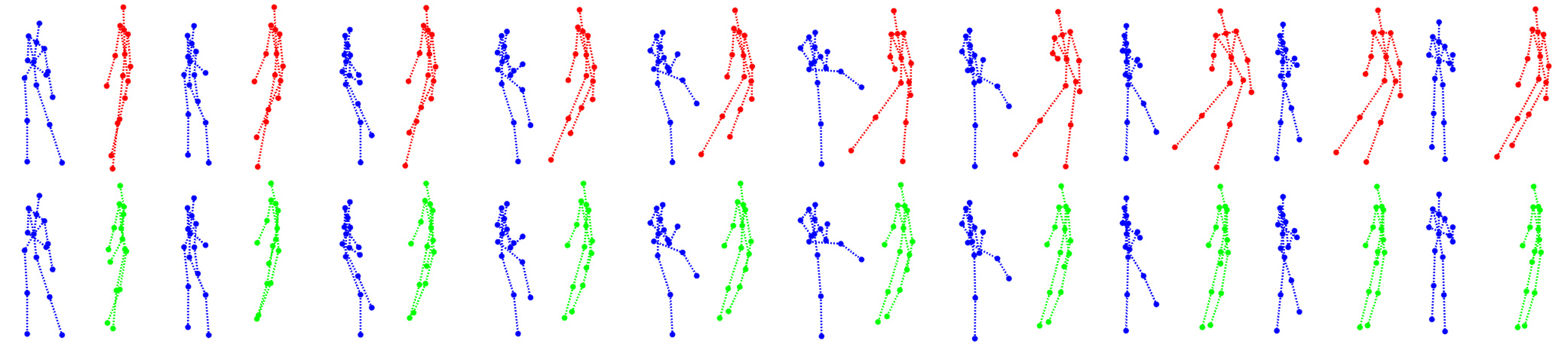}}
    \subfigure[Push]{
        \includegraphics[width=0.8\linewidth]{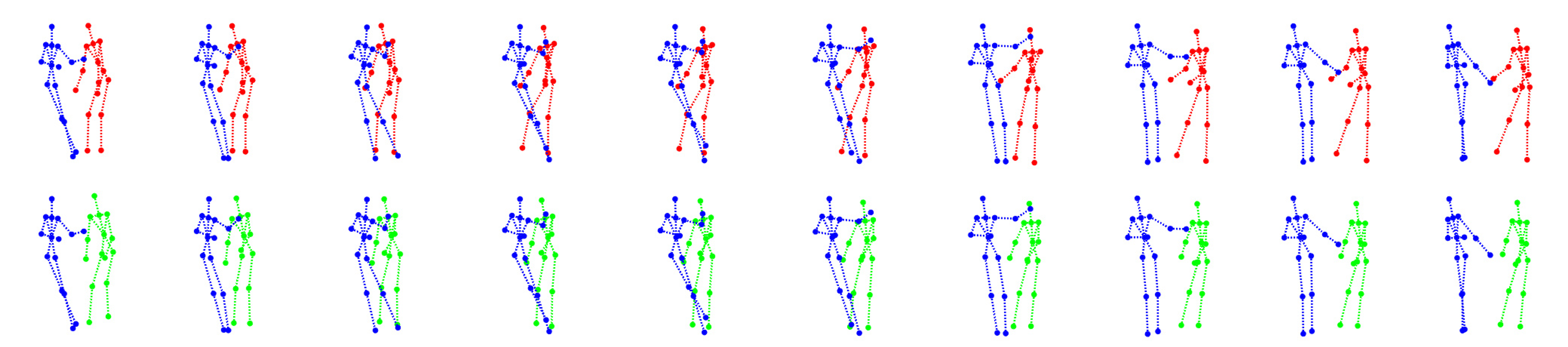}}
    \subfigure[Shake Hands]{
        \includegraphics[width=0.8\linewidth]{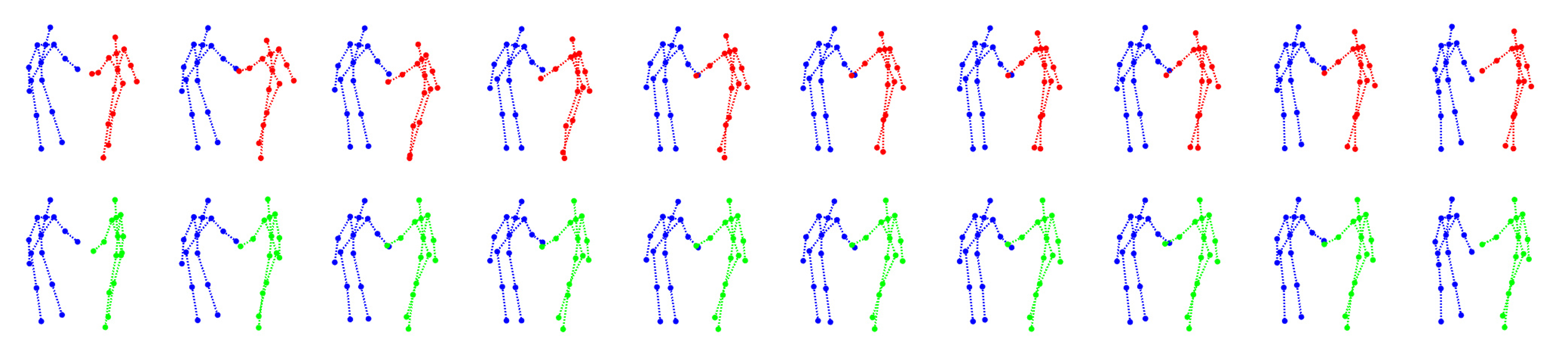}}
    \subfigure[Hug]{
        \includegraphics[width=0.8\linewidth]{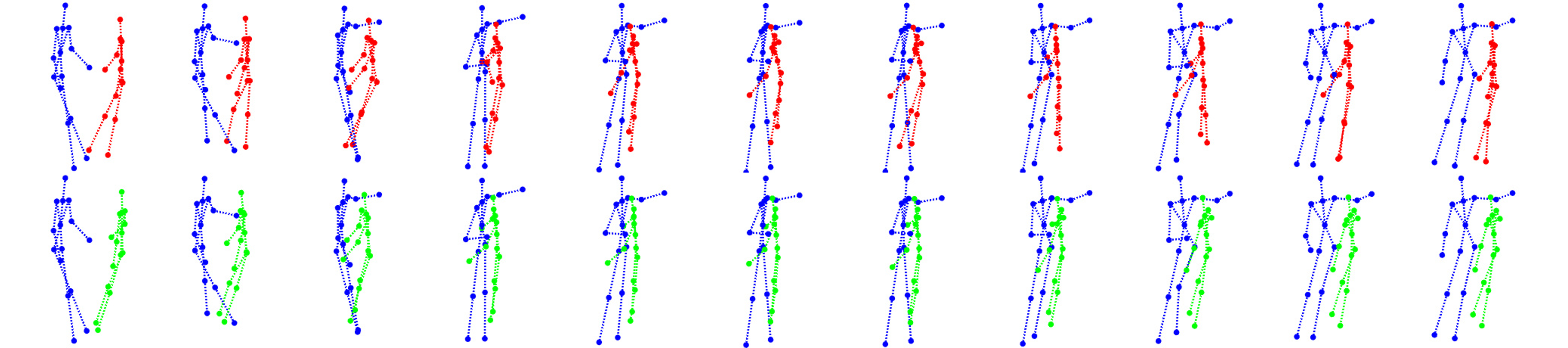}}
    \subfigure[Exchange Objects]{
        \includegraphics[width=0.8\linewidth]{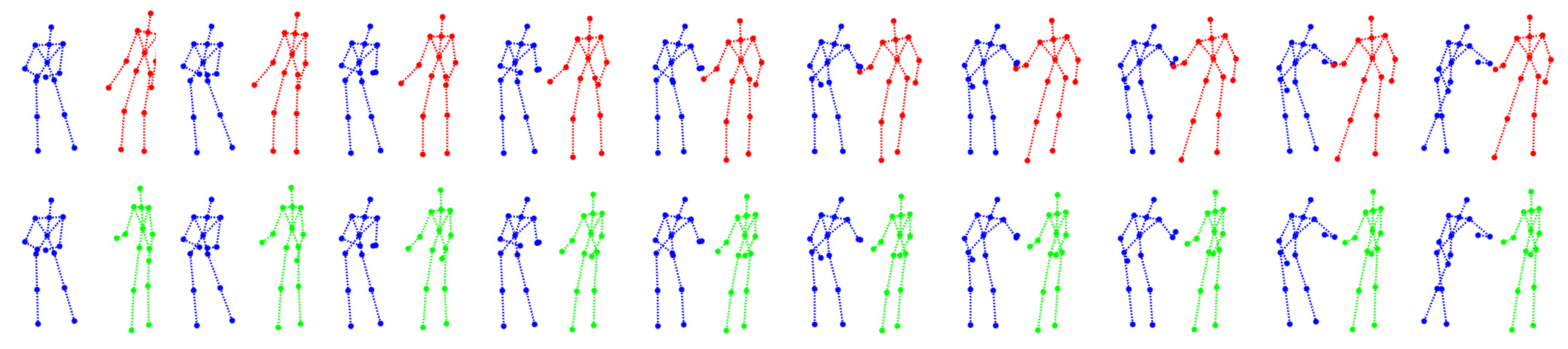}}
    \subfigure[Punch]{
        \includegraphics[width=0.8\linewidth]{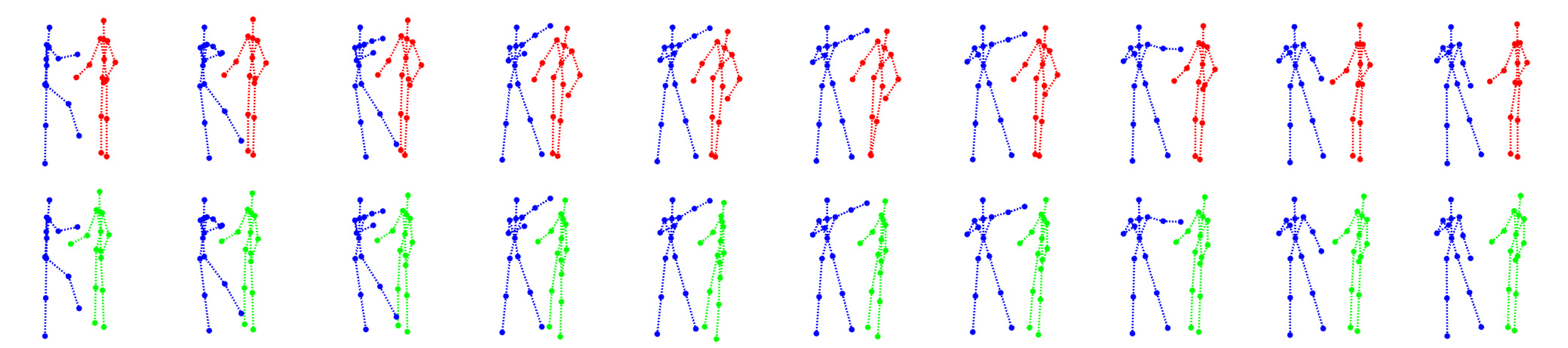}}
    \caption{Examples of real and synthesized reactive motion of the six interaction classes in the SBU dataset. Given the input motion (blue) and the interaction label, the reactive motion (green) can be synthesized. The ground truth reactive motion (red) is also included for comparison.}
    \label{fig:sbu_multi}
\end{figure}

\paragraph{Interaction Match - Comparing with Ground Truth Data} We first demonstrate the resemblance of data from the SBU dataset and the results are illustrated in Figure \ref{fig:sbu_multi}. Specifically, we employ a leave-one-subject-out cross-validation approach to train our model. At the inference stage, the input motion (colored in blue) alongside the ground truth interaction label is used for generating the reactive motion (colored in green). It can be seen that the synthesized reactive motions resemble the corresponding interactions as in the ground truth data (colored in red). Since our proposed method is based on a generative model, some variations are introduced to the reactive motions when compared with the ground truth data. For example, the leg movements of \textit{Kick} (Figure \ref{fig:sbu_multi}(a)) and \textit{Exchange Objects} (Figure \ref{fig:sbu_multi}(e)) are different from their corresponding ground truth motions. Nevertheless, the context of the interactions is correctly preserved.


\begin{figure*} [htb]
\centering
    \subfigure[Kick Sample One]{
        \includegraphics[width=0.8\linewidth]{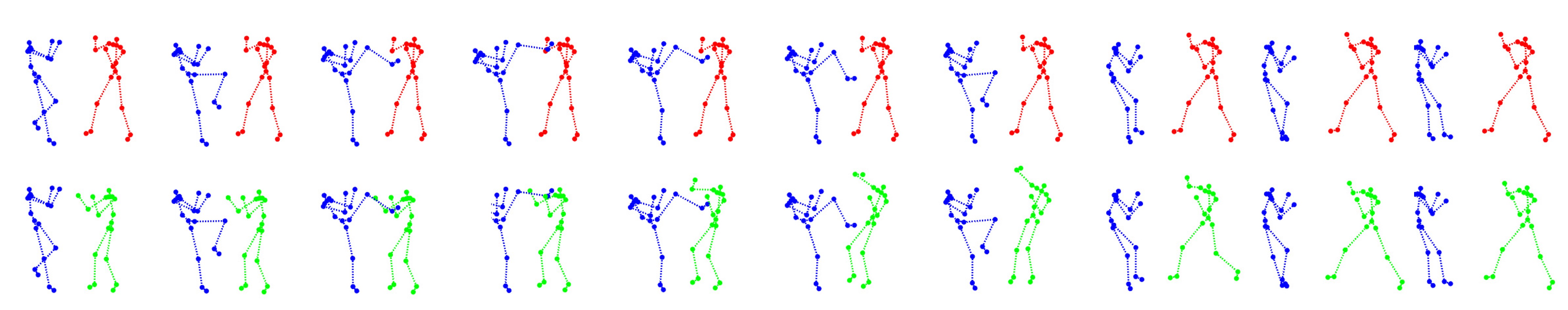}}
    \subfigure[Kick Sample Two]{
        \includegraphics[width=0.8\linewidth]{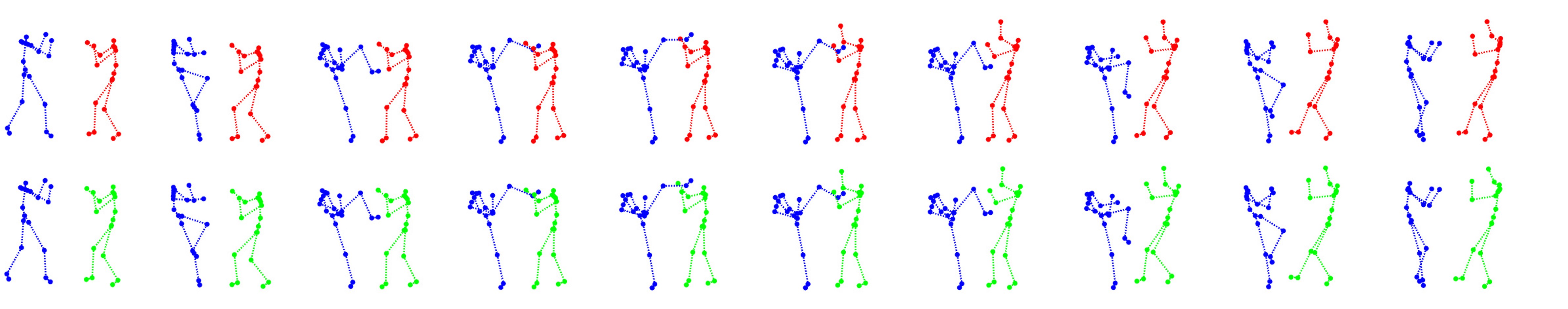}}
    \subfigure[Punch Sample One]{
        \includegraphics[width=0.8\linewidth]{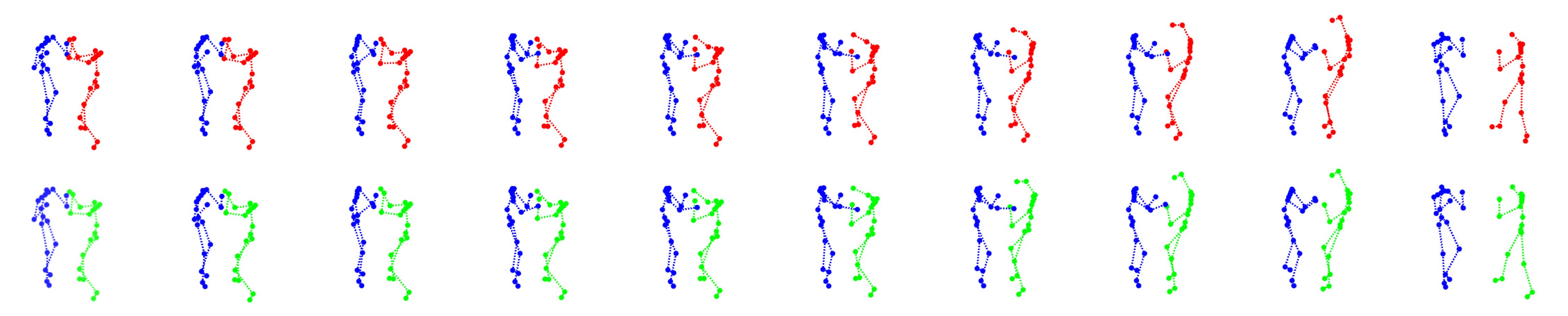}}
    \subfigure[Punch Sample Two]{
        \includegraphics[width=0.8\linewidth]{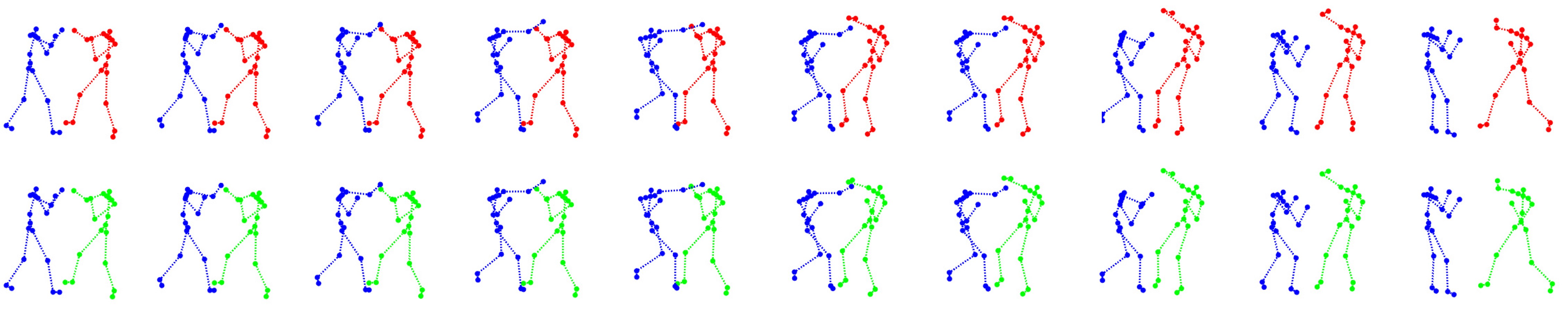}}
    \caption{Examples of real and synthesized reactive motion of kicking and punching interaction in the 2C dataset, respectively. Given the input motion (blue) and the interaction label, the reactive motion (green) can be synthesized. The ground truth reactive motion (red) is also included for comparison.}
    \label{fig:2c}
\end{figure*}

We further demonstrate the capability and the generality of our proposed method by generating reactive motions on the high-quality 2C dataset. Examples of the synthesized motions are illustrated in Figure \ref{fig:2c}. Similar to the results obtained in the SBU dataset, our method can resemble the interactions in the 2C dataset while introducing some variations to the synthesized reactive motions.

\begin{figure*}[htb]
   \centering
    \subfigure[Shaking hands and hugging (green) while kicking (blue)\new{, with the Multi-hot labels \textit{Shaking(+1)} and \textit{Hugging(+1)}}]{
        \includegraphics[width=0.9\linewidth]{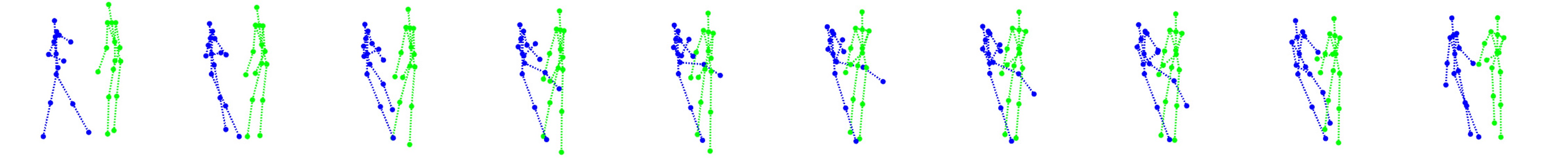}}
    \subfigure[Getting pushed and shaking hands (green) while pushing (blue)\new{, with the Multi-hot labels \textit{Pushed(+1)} and \textit{Shaking Hands(+1)}}]{
        \includegraphics[width=0.9\linewidth]{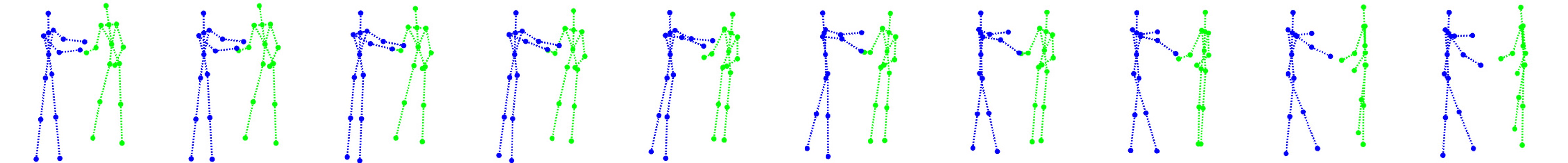}}
    \subfigure[Getting hit and pushed back (green) while kicking (blue)\new{, with the Multi-hot label \textit{Kicking(-1)}}]{
        \includegraphics[width=0.9\linewidth]{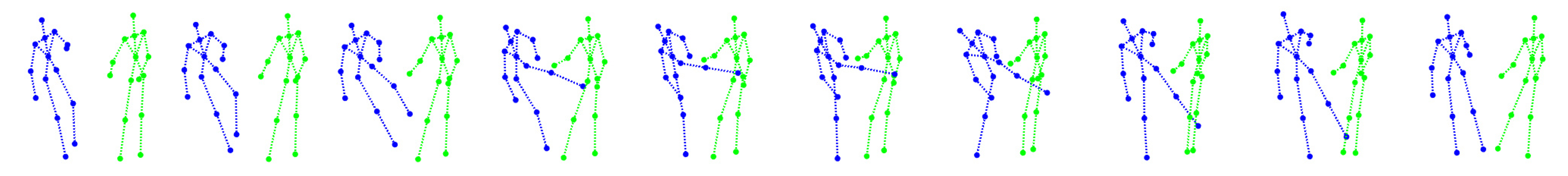}}
    \subfigure[Getting hit and pushed back (green) while punching (blue)\new{, with the Multi-hot label \textit{Punching(-1)}}]{
        \includegraphics[width=0.9\linewidth]{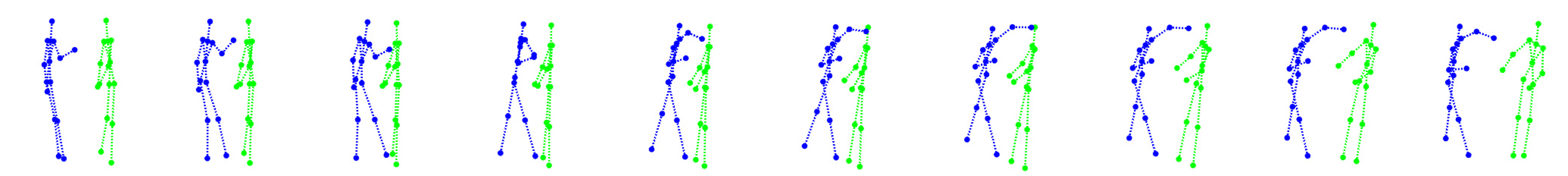}}
   \caption{\new{Examples of Interaction Mix (a and b) and Match (c and d) on the SBU dataset along with their Multi-hot Labels. The input and the synthesized reaction are colored blue and green, respectively.}}
   \label{fig:kick_mm2}
\end{figure*}

\paragraph{Interaction Mix}
With the multi-hot embedding in our proposed framework, users can control the reactive motions to be synthesized by specifying the interaction labels as a multi-hot vector. Some examples generated from the SBU dataset are shown in Figure \ref{fig:kick_mm2}. It can be seen that the generations show hybrid reactive patterns. For example, we set the labels of shaking hands and hugging to be positive to generate the reaction in (a). We observe that the synthesized character B is getting close to character A meanwhile ready to shake hands (as shown in the last few frame stamps). In (b), the generated motion for character B is shaking hands with A while laying back to avoid being pushed.

\subsection{Analysis of the Latent Space} \label{sec:exp_latent}
\begin{figure*}[htb]
  \includegraphics[width=1\linewidth]{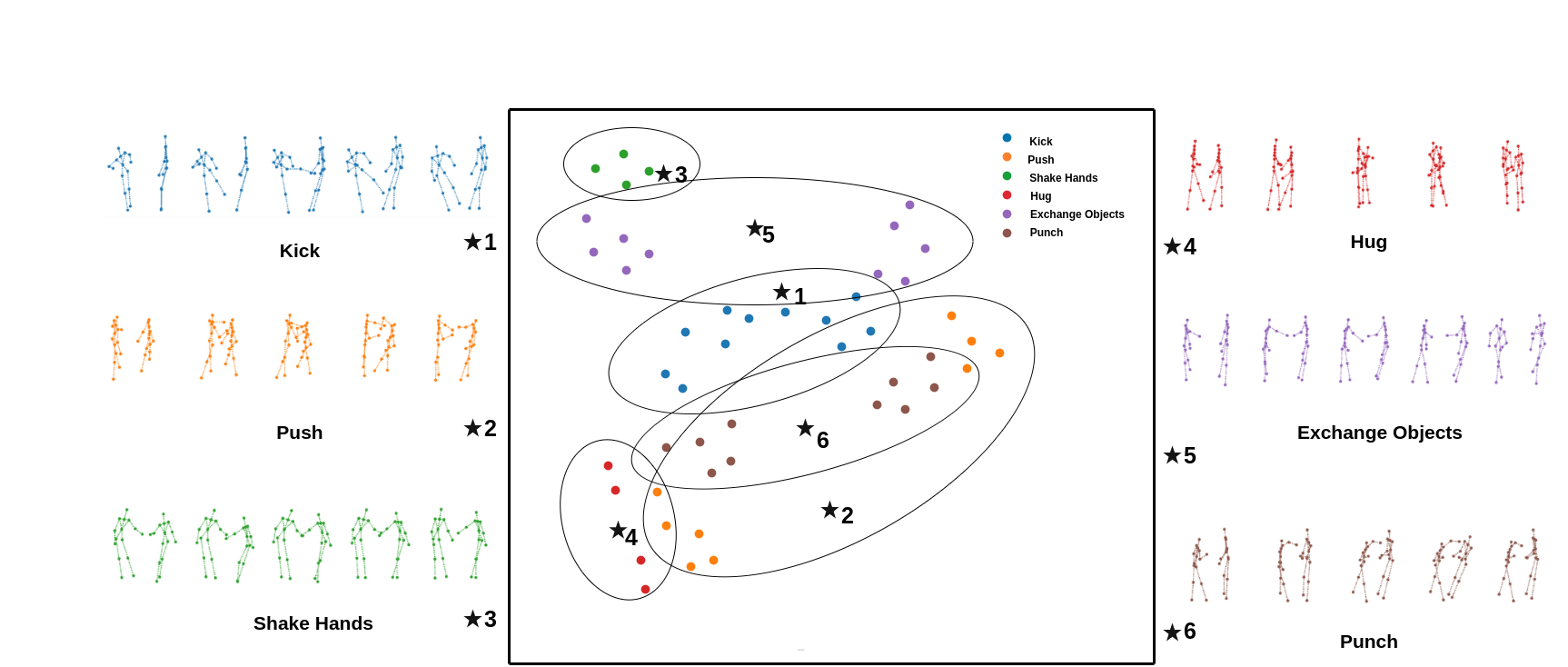}
  \caption{t-SNE of the embedding space obtained using our proposed method. }
  \label{fig:t-SNE}
\end{figure*}

To analyze the quality of the latent space learned, the t-SNE of the embedding space is illustrated in Figure \ref{fig:t-SNE}. Specifically, we take the embedding of the reactive interaction generated, just after the concatenation of the Conditional Hierarchical Bi-LSTM outputs and run t-SNE (t-distributed stochastic neighbor embedding). Again, we follow the protocol to have a leave-one-subject-out cross-validation in this experiment.  

The results indicate that most of the interaction classes formed clusters which show a low degree of inter-class similarity. In particular, 4 out of 6 classes, including \textit{Exchange Objects, Shake Hands, Hug} and \textit{Kick} are not overlapping with the other classes. Although there is an overlapping between the regions covered by the clusters of the \textit{Punch} and \textit{Push} classes, it can be seen that the movements of the character who punches or pushes the other are very similar in terms of the skeletal motions. The corresponding reactive motions are also similar in those two classes. Furthermore, the samples of the \textit{Push} are essentially outside the cluster of \textit{Punch}. This highlight the effectiveness of the learning of the latent space for representing different types of interactions. In addition to separating different types of interactions, Figure \ref{fig:t-SNE} also highlights our latent space can effectively capture the similarity between different classes. For example, \textit{Shake Hand} and \textit{Exchange Object} are more similar as well as the \textit{Punch} and \textit{Push} pair discussed above.

\begin{figure}[htb]
   \centering
  \includegraphics[width=0.5\linewidth]{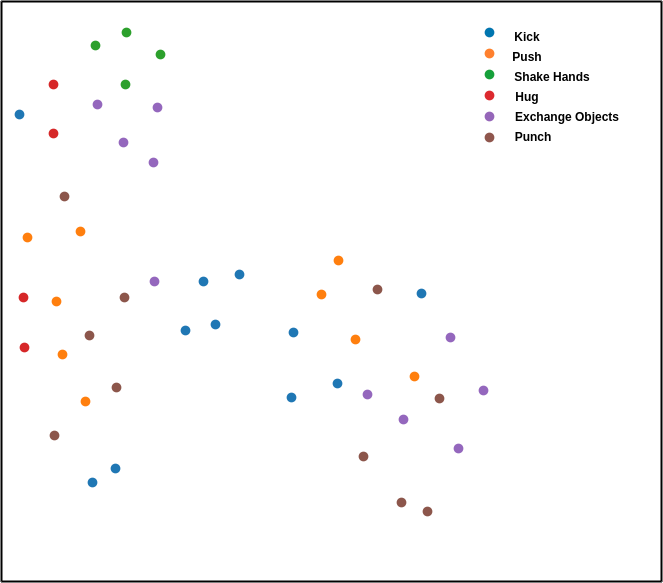}
  \caption{t-SNE of the embedding space obtained using \cite{Men:CG2022}. }
  \label{fig:t-SNE_CnG}
\end{figure}

We further compare the t-SNE of the embedding space obtained using \cite{Men:CG2022} in Figure \ref{fig:t-SNE_CnG}. It can be seen that samples from different interaction types are mixed together in the latent space. We argue that a more well-constructed latent space not only informing the user about which interaction types are more suitable for \textit{Interaction Mix} based on their similarity, but also facilitates the extrapolation of different types of interaction with the multi-hot label in \textit{Interaction Mix}.

\subsection{Ablation Study} \label{sec:exp_ablation}
In this section, we justify the design of different components in our proposed framework by conducting an ablation study. \new{The results are presented in Table \ref{tab:ablation_AFD} and \ref{tab:ablation_AFD_2C} for the SBU and 2C datasets, respectively.} 

From the results in Table \ref{tab:ablation_AFD}, our proposed network without the bone loss ($\mathcal{L}_{b}$) achieved the lowest AFD across all 6 classes. The complete version of our framework achieved the lowest AFD in the \textit{Punch} class and second-lowest in the rest of the 5 classes. Since the SBU dataset is captured using Microsoft Kinect, we observed that the skeleton motions captured are not of high-quality and the bone lengths of the characters vary over time. As a result, having the bone loss maintains the bone lengths over time will not give our method a favourable AFD on this dataset, although it is essential to enforce such a constraint in rigid-body character animation.   

\new{We also show the ablation test on the high-quality 2C dataset with the results given in Table \ref{tab:ablation_AFD_2C}. By combining all constraints, our model shows the best motion quality with the lowest FID. Since the bone length captured in 2C is more stable, the advantage of continuity constraint $\mathcal{L}_c$ is more advantageous than $\mathcal{L}_b$ compared to SBU dataset. Furthermore, the effectiveness of the multi-hot encoding is also significant when generating between these two highly-similar interaction.}

\begin{table}[htb]
    \centering
\begin{tabular}{|l|c|c|c|c|c|c| } 
 \hline
 & \multicolumn{6}{c|}{AFD (↓)} \\
 \hline
 Method & Kick & Push & Shake hand & Hug & Ex. Obj. & Punch \\
 \hline
OURS   & 0.5   & 0.43 & 0.4 & 0.42 & 0.4 & \textbf{0.32}\\ \hline
OURS w/o $D$   & 0.53    & 0.45 & 0.42 & 0.49 & 0.44 &  0.38\\
OURS w/o $\mathcal{L}_{b}$ & \textbf{0.45} & \textbf{0.42} & \textbf{0.39} & \textbf{0.40} & \textbf{0.39} & \textbf{0.32}\\
OURS w/o $\mathcal{L}_{c}$ & 0.59 & 0.45 & 0.55 & 0.46 & 0.41 & 0.4\\
OURS w/o FC & 0.55 & 0.47 & 0.46 & 0.49 & 0.41 & 0.4\\
OURS w/o Multi-Hot Embed & 0.51 & 0.48 & 0.50 & 0.66 & 0.46 & 0.45\\
 \hline
\end{tabular}%
    \caption{Ablations of main components in the proposed model evaluated by AFD on the SBU dataset. \new{The model without the bone loss ($\mathcal{L}_{b}$) gives the lowest AFD}}
    \label{tab:ablation_AFD}
\end{table}

\begin{table}[htb]
    \centering
\begin{tabular}{|l|c|c| } 
 \hline
 & \multicolumn{2}{c|}{FID (↓)} \\
 \hline
Method  & Kick & Punch  \\
 \hline
OURS   & \textbf{164.2}   &  \textbf{122.4} \\ \hline
OURS w/o $D$ & 174.4  & 128.2 \\
OURS w/o $\mathcal{L}_{b}$ & 166.4 & 124.8\\
OURS w/o $\mathcal{L}_{c}$ & 180.2 & 131.4\\
OURS w/o FC & 175.3 & 129.1 \\
OURS w/o Multi-Hot Embed & 175.1 & 129.7 \\
 \hline
\end{tabular}
    \caption{Ablations of main components in the proposed model evaluated by FID on the 2C dataset. \new{Our model with all loss terms shows the lowest FID}}. 
    \label{tab:ablation_AFD_2C}
\end{table}



\subsection{Data Augmentation for Interaction Recognition} \label{sec:recognitoin}
To demonstrate another application of our proposed method, we conduct a series of experiments to evaluate how the interactions synthesized by our method can be used for enhancing the datasets for training interaction recognition algorithms. Specifically, we generate a new set of 300 (50 per class) interactions from the SBU dataset and the dataset will be available to the public to stimulate the research in this area. We train an interaction classifier to test the quality of the synthesized interaction. The classifier has the same structure as the multi-class discriminator but outputs N classes instead of N+1. Its train-test split settings are as follows:
\begin{itemize}
    \item \textbf{Original}: We follow the half-half data split \cite{Yun:SBU} widely used in the interaction recognition protocol in the SBU dataset
    \item \textbf{Augmented}: We evenly divided the synthesized dataset and add them to the original SBU training and testing set 
\end{itemize}

The classification accuracies are reported in Table \ref{tab:recognition}. The results show that the variations introduced by our augmented dataset have a positive impact on improving the interaction classification performance over the original SBU dataset. In particular, 5 out of 6 classes have an increase in the classification accuracy with a range from 1.33\% to 21.39\%. 


\begin{table*}[htb]
    \centering
\begin{tabular}
{ |l|c|c||c|c|c|c|c|c| }
 \hline
 \multicolumn{3}{|c||}{} & \multicolumn{6}{c|}{Accuracy} \\
 \hline
 Different Splits & \# Train Seq. & \# Test Seq. & Kick & Push & Shake Hand & Hug & Exchange Object & Punch \\
 \hline
Original& 99 & 98   & \textbf{0.9474}    &0.9411 & 0.6923 & 0.5385 & 0.6674 & 0.8889 \\
Augmented & 124 & 123& \textbf{0.9960} & 0.8770 & \textbf{0.7056} & \textbf{0.6407} & \textbf{0.8813} & \textbf{0.9112} \\

 \hline
\end{tabular}
\caption{Accuracy of the classifier on different splits on the original and augmented SBU datasets. \new{The augmented dataset helps improve the recognition performance over most of the interaction classes in the original SBU dataset.}}
    \label{tab:recognition}
\end{table*}

\section{Discussion and Conclusions}

In this paper, we propose a novel GAN-based framework for synthesizing 2-character interactions based on the input motion of one character and the interaction label. By incorporating the multi-hot class embedding, our method enables \textit{Interaction Mix} which generates new interaction classes by extrapolating single-hot labelled interactions for the training data, as well as \textit{Interaction Match} which synthesizes diverse interaction variations from the original interaction classes in the dataset. Experimental results show that our method outperforms the existing work including the most relevant work~\cite{Men:CG2022}.

\begin{figure}[htb]
   \centering
    \includegraphics[width=0.7\linewidth]{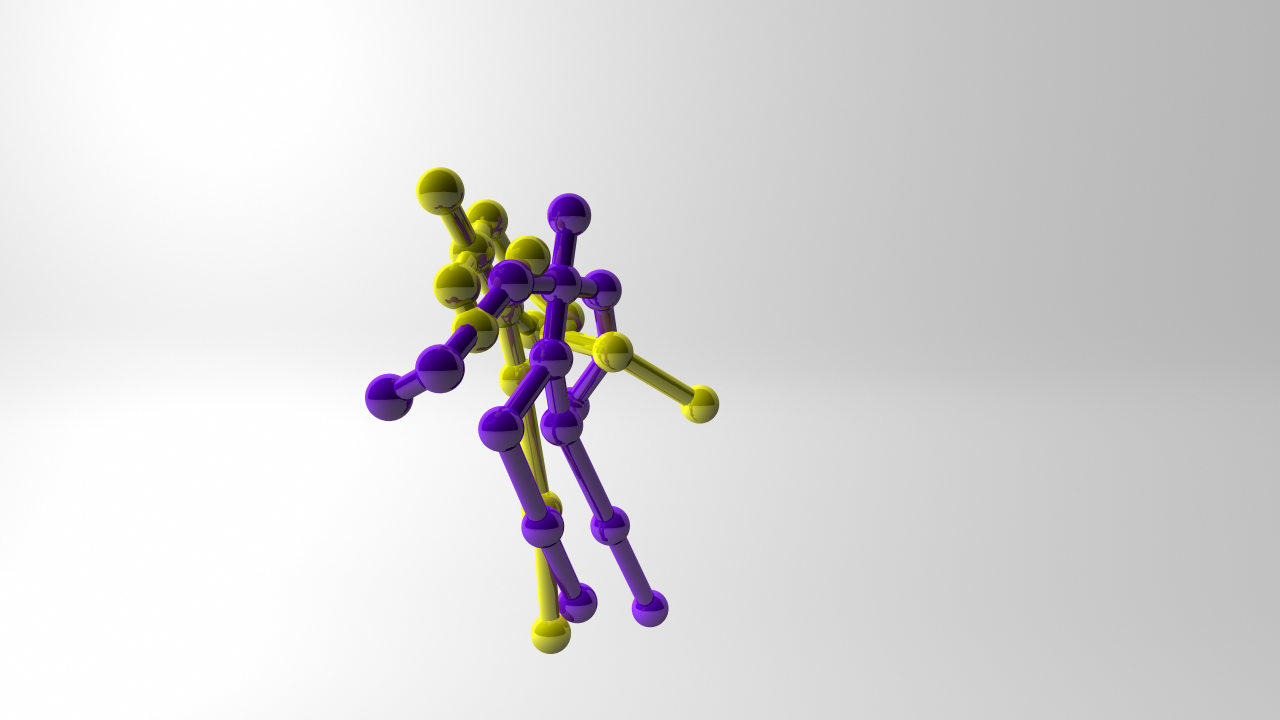}
   \caption{Artefacts such as interpenetration of body parts can be found when mixing very dissimilar interaction types, e.g. mixing Kicking and Hugging. }
   \label{fig:failure}
\end{figure}

\newedit{While the proposed method provides users with a large degree of high-level control, mixing very dissimilar interaction types may result in artefacts such as interpenetration of body parts (see the example illustrated in Figure \ref{fig:failure})}. This can happen since our method does not handle collisions explicitly. Further exploring the usage of the latent space, such as interpolation and extrapolation, learned using our model as well as learning a topology-aware latent space \cite{Ho:TopologyIK:PG13} for avoiding interpenetration are potential future directions. \new{Using existing close interaction editing methods such as \textit{Interaction Mesh}~\cite{Ho:ToG10} and \textit{Aura Mesh}~\cite{Jin:CFG18} as a post-processing step to clean up the interpenetration as well as maintain the contact points between the characters can be another solution as demonstrated in \cite{Ho:TOMM13}. In terms of the quality of the synthesized motion, artefacts such as foot sliding can be found in the synthesized motions since there is no explicit loss term on the stepping pattern in our proposed network and this is quite common to other GAN-based motion synthesis methods \cite{Aristidou:TVCG2022} such as MotionCLIP~\cite{MotionCLIP}. We are interested in incorporating the contact consistency loss proposed in GANimator~\cite{li2022ganimator} or an additional post-processing step \cite{He:TVCG2021} to clean up the foot sliding artefacts in the results as an extension of the proposed work. }

In the future, we could use a classifier\cite{7926607} to generate the Multi-Hot Class Embedding during training and inference time. The Multi-Hot Class Embedding could also be extended to single body motion synthesis~\cite{Battan_2021_WACV} to generate controllable synthesized motions.
\bibliographystyle{unsrt} 
\bibliography{interaction}        



\end{document}